\documentclass[graybox]{svmult}

\usepackage{mathptmx}       % selects Times Roman as basic font
\usepackage{helvet}         % selects Helvetica as sans-serif font
\usepackage{courier}        % selects Courier as typewriter font
\usepackage{type1cm}        % activate if the above 3 fonts are
\usepackage{makeidx}         % allows index generation
\usepackage{graphicx}        % standard LaTeX graphics tool
\usepackage{multicol}        % used for the two-column index
\usepackage[bottom]{footmisc}% places footnotes at page bottom

\makeindex             % used for the subject index
\begin{document}

\title*{The complex interplay of dust and star light in spiral galaxy discs}
\author{Maarten Baes, Dimitri Gadotti, Joris Verstappen, Ilse De
  Looze, Jacopo Fritz, Edgardo Vidal P\'erez and Marko Stalevski}
\authorrunning{Maarten Baes et al.}
\institute{Maarten Baes, Joris Verstappen, Ilse De Looze, Jacopo
  Fritz, Edgardo Vidal P\'erez \at Sterrenkundig Observatorium, Universiteit
  Gent, Krijgslaan 281 S9, B-9000 Gent, Belgium,
  \email{maarten.baes@ugent.be}
\and Dimitri Gadotti \at European Southern Observatory, Alonso de
Cordova 3107, Vitacura, Santiago, Chile
\and Marko Stalevski \at Belgrade Astronomical Observatory, Volgina 7,
P.O. Box 74, 11060 Belgrade, Serbia}
\maketitle

\abstract{Interstellar dust grains efficiently absorb and scatter UV
  and optical radiation in galaxies, and therefore can significantly
  affect the apparent structure of spiral galaxies. We discuss the
  effect of dust attenuation on the observed structural properties of
  bulges and discs. We also present some first results on modelling
  the dust content of edge-on spiral galaxies using both optical and
  Herschel far-infrared data. Both of these results demonstrate the
  complex interplay of dust and star light in spiral galaxies. }

\section{Introduction}

In has been known for a long time that interstellar dust grains are an
important component of the interstellar medium in galaxies: they
efficiently absorb and scatter UV and optical radiation, play an
important role in the chemistry of the ISM and are the dominant source
of far-infrared and submillimetre emission. Detailed knowledge of the
quantity, spatial distribution and physical properties of the dust in
spiral galaxies is still controversial in many ways. For several
decades after the work by Holmberg (1958), it was assumed that spiral
galaxies were optically thin for optical radiation. This conventional
viewpoint was questioned in the late 1980s and early 1990s by several
teams, including Disney et al.\ (1989), Valentijn (1990) and Burstein
et al.\ (1991). These authors came to the conclusion that spiral
galaxies are optically thick, even in the outer regions. Around the
same time, other teams reached completely different conclusions,
sometimes on the basis of identical data sets (e.g.\ Huizinga \& van
Albada 1992).

In retrospective, three issues conspired to complicate the discussion
on the optical thickness of spiral galaxies. The first was
observational biases in the classical tests, such as the expected
variation of isophotal diameter, mean surface brightness with
increasing inclination. It is very difficult to quantify these biases,
let alone to circumvent them (e.g.\ Davies et al.\ 1993). A second
issue that complicated the discussion was the unavailability of
reliable far-infrared observations that traced the bulk of the dust
mass in spiral galaxies. In particular, almost no far-infrared or
submillimetre data were available redwards of 100~$\mu$m, the longest
IRAS wavelength. The final and possibly the most important reason why
the contrary results were obtained was the simplicity of the models
used to analyze or interpret the data. This was convincingly
demonstrated by Disney et al.\ (1989), who showed that simple
optically thick models could reproduce the same observations on the
basis of which Holmberg (1958) had reached his conclusions.

One of the first to realize that realistic models for dusty galaxies
were absolutely needed to investigate the dust content of spiral
galaxies was Ken Freeman. He and his Ph.D. student Yong-Ik Byun
embarked on the first systematic effort to investigate the complex
interplay of dust and star light in realistic models of spiral
galaxies, using detailed radiation transfer simulations including both
absorption and scattering. The result of this seminal work, presented
in Byun et al.\ (1994), is still a monument in the extragalactic
radiative transfer community. 

In the past 15 years, the modelling of the interplay between dust and
star light in spiral galaxies has been dealt with by many authors.  In
particular, several groups have developed the necessary numerical
codes to solve the radiative transfer problem in extragalactic
environments. The most advanced of these codes are not restricted to
only absorption and scattering in 2D geometries, but take into account
thermal emission by dust, polarization, kinematics and multi-phase
dust distributions (e.g.\ Popescu et al.\ 2000; Gordon et al. 2001;
Misselt et al.\ 2001; Steinacker et al.\ 2003; Bianchi 2008). Our
group has also developed a 3D radiative transfer code, SKIRT, based on
the Monte Carlo method. Its original aim was to investigate the
effects of dust absorption and scattering on the observed kinematics
of elliptical galaxies (Baes \& Dejonghe 2000, 2002; Baes et al.\
2000). It has now developed into a mature radiative transfer code that
can be used to simulate images, spectral energy distributions,
kinematics and temperature maps of dusty systems, ranging from
circumstellar discs to AGNs (e.g.\ Vidal et al.\ 2007; Stalevski et
al.\ 2010). In particular, the code has been used as the main tool for
a detailed investigation of radiative transfer in spiral galaxies
(Baes \& Dejonghe 2001a, b; Baes et al.\ 2003). In the remainder of
this paper we shortly describe two recent results that demonstrate the
complex interplay of dust and star light in spiral galaxies: in
Section~2 we discuss the effect of dust attenuation on the observed
structural properties of bulges and discs, and in Section~3 we present
some first results on modelling the dust content of edge-on spiral
galaxies using both optical and far-infrared data.

\section{Dust effects on bulge and disc parameters}

It has been known for many decades that the presence of dust
influences the observed, apparent photometric galaxy parameters
(apparent scalelengths, surface brightnesses, luminosities, axial
ratios, etc.) and makes it a challenge to recover the intrinsic
unaffected parameters. Several authors have investigated these effects
using radiative transfer modelling with varying degrees of
sophistication and/or geometrical realism. In general, it was found
that the importance of dust attenuation varies as a function of
wavelength, galaxy inclination and star-dust geometry. In particular,
Byun et al.\ (1994) were the first to convincingly demonstrate that
effects of scattering are often counter-intuitive and crucial to
properly interpret the effects of dust.

In the last few years, two independent teams have investigated the
effects of dust attenuation in bulge and disc components, on their
integrated properties, separately, using realistic models of spiral
galaxies. Both Pierini et al.\ (2004) and Tuffs et al.\ (2004)
presented attenuation functions for the individual disc and bulge
components of dusty spiral galaxies. They clearly demonstrated that
the effects of dust on the bulge and disc components can differ
substantially, as a result of the different star-dust geometry. We
have aimed to extend this work one stage further. We have embarked on
a project to investigate the systematic effects of dust attenuation on
the apparent detailed structural properties of discs and bulges
simultaneously. We have created artificial galaxy images, using
radiative transfer simulations, to mimic the observed structural
properties of disc galaxies with classical and pseudo-bulges, and
include the effects of dust attenuation in the observed light
distribution. By applying 2D bulge/disc decomposition techniques in
this set of models, we were able to evaluate what are the effects of
galaxy inclination and dust opacity on the results from such
decompositions.

\begin{figure}[t!]
  \centering
 \includegraphics[angle=-90,width=0.5\textwidth]{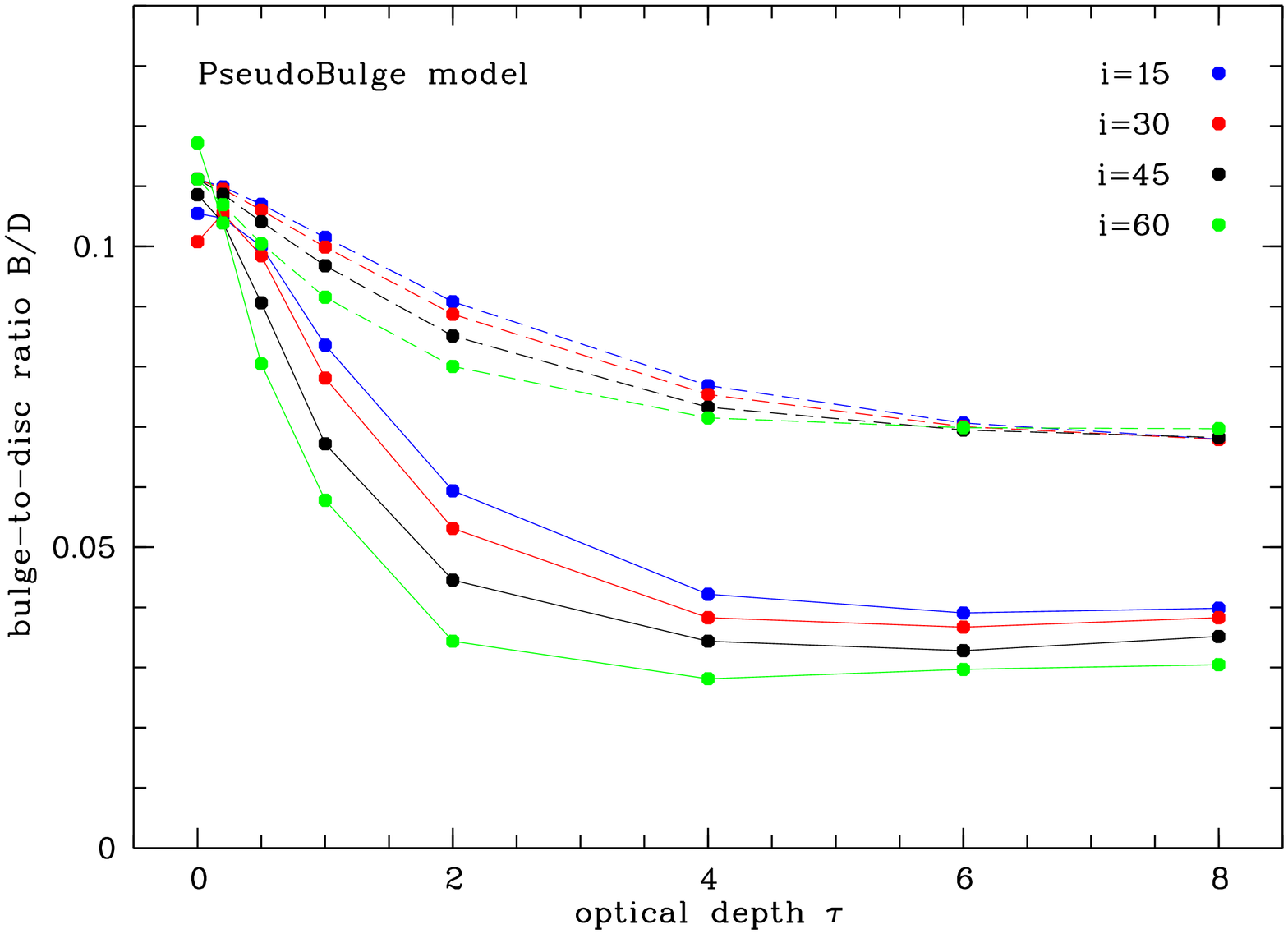}%
  \includegraphics[angle=-90,width=0.5\textwidth]{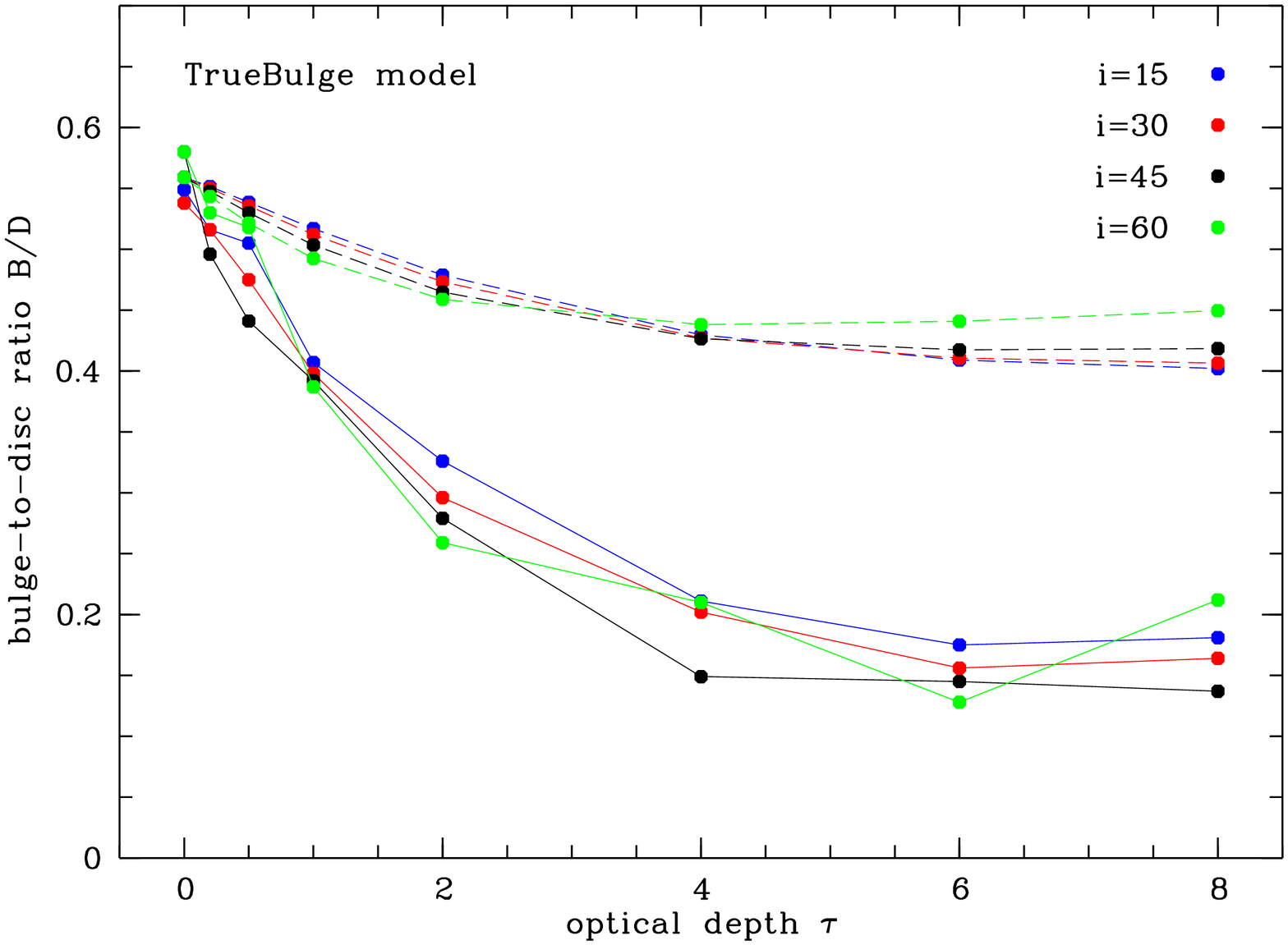}
  \caption{Dependence of the bulge-to-disc ratio on the V-band optical
    depth $\tau$. The solid lines represent the apparent bulge-to-disc
    ratio as derived from the {\sc budda} bulge/disc decompositions of
    the dust-affected images. The dashed lines represent the actual
    bulge-to-disc ratio as determined from the ratio of the input
    bulge and disc integrated fluxes.}
  \label{BulgeToDiscRatio.pdf}
\end{figure}

Rather surprisingly, we have found that the effects of dust on the
structural parameters of bulges and discs obtained from 2D bulge/disc
decomposition cannot be simply evaluated by putting together the
effects of dust on the properties of bulges and discs treated
separately. In particular, the effects of dust in galaxies hosting
pseudo-bulges might be different from those in galaxies hosting
classical bulges, even if their dust content is identical. Confirming
previous results, we find that disc scalelengths are overestimated
when dust effects are important. In addition, we also find that bulge
effective radii and S\'ersic indices are underestimated. Furthermore,
the apparent attenuation of the integrated disc light is
underestimated, whereas the corresponding attenuation of bulge light
is overestimated. Dust effects are more significant for the bulge
parameters, and, combined, they lead to a strong underestimation of
the bulge-to-disc ratio, which can reach a factor of 2 in the V band,
even at relatively low galaxy inclinations and dust opacities (see
Fig.~{\ref{BulgeToDiscRatio.pdf}}). The reason for these, at first
sight, counter-intuitive results comes from the fact that such
decompositions use specific models to fit bulges and discs which
cannot accommodate the effects of a dust disc in the
galaxy. Therefore, when the model for a component tries to adjust
itself when dust is present, this has direct consequences on the model
of the other component, even if the latter is not directly affected by
dust. More details can be found in Gadotti et al.\ (2010).

\section{The energy balance in spiral galaxies}

Apart from more advanced modelling techniques, one of the crucial
steps forwards (compared to the early 1990s) in determining the dust
content of spiral galaxies is the availability of observations in the
far-infrared range between 100 and 1000 $\mu$m. Observations in this
wavelength range are crucial because they directly trace the smoothly
distributed cold dust in spiral galaxies, which is heated to
temperatures of some 20~K by the general interstellar radiation
field. While the ISO and Spitzer missions have opened up the
far-infrared window out to 200~$\mu$m, the most important step forward
was the launch of Herschel in May 2009. The first results of Herschel
imaging of nearby galaxies such as M81, M82 or NGC\,4438 are
spectacular (e.g.\ Bendo et al.\ 2010; Roussel et al.\ 2010; Cortese
et al.\ 2010).

We are focusing our attention to edge-on spiral galaxies, because they
are an important class of galaxies in which the distribution and
properties of interstellar dust grains can be studied in great
detail. On the one hand, the dust in these systems shows prominently
as dust lanes in optical images; on the other hand, surface brightness
enhancements ensure that the far-infrared dust emission in edge-on
spirals can be traced out to large radii. A self-consistent treatment of
extinction and thermal emission, i.e. a study of the dust energy
balance, gives the strongest constraints on the dust content of spiral
galaxies.  

For several edge-on spiral galaxies, the dust distribution has been
modeled by fitting realistic radiative transfer models to such optical
images (Kylafis \& Bahcall 1987; Xilouris et al. 1997, 1998, 1999;
Alton et al.  2004; Bianchi 2007). The conclusion of these works is
that, in general, the dust disc is thinner (vertically) but radially
more extended than the stellar disc and that the central optical depth
perpendicular to the disc is less than one in optical wavebands,
making the disc almost transparent when seen face-on. This result
seems to be at odds with FIR/submm emission studies, which indicate
that spiral galaxies typically reprocess about 30\% of the UV/optical
radiation (Popescu \& Tuffs 2002). When applied to individual edge-on
spiral galaxies, it is found that the predicted FIR fluxes of
self-consistent radiative transfer models that successfully explain
the optical extinction generally underestimate the observed FIR fluxes
by a factor of about three (Popescu et al. 2000; Misiriotis et
al. 2001; Alton et al. 2004; Dasyra et al. 2005).  Several scenarios
have been proposed to explain this discrepancy, but a major problem
discriminating between these is that the number of edge-on galaxies
for which such detailed studies have been done so far is limited, due
to the poor sensivity, spatial resolution and limited wavelength
coverage of the available FIR instruments. 

\begin{figure}[t!]
\sidecaption[t]
\includegraphics[width=0.6\textwidth]{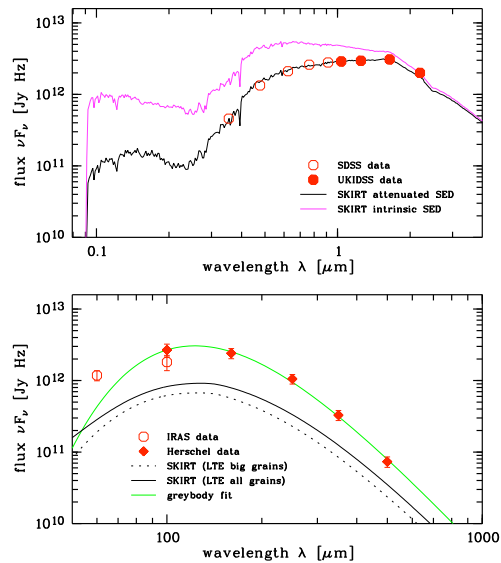}
\caption{The optical/NIR (top) and FIR/submm (bottom) spectral energy
  distribution of UGC\,4754. The solid black line in the top panel
  corresponds to the attenuated SED of the SKIRT model fitted to the
  SDSS and UKIDSS images, the magenta line is the unattenuated SED. In
  the bottom panel, the solid black line corresponds to the FIR
  emission of the model assuming LTE for all grains; the dotted line
  represents the contribution of the large grains only. Our model
  significantly underestimates the observed IRAS and Herschel
  fluxes. }
\label{SED.pdf}
\end{figure}

We have used Herschel PACS and SPIRE observations of the edge-on
spiral galaxy UGC\,4754 to investigate its dust energy balance (for
more details, see Baes et al.\ 2010). We build detailed SKIRT
radiative models based on SDSS and UKIDSS maps, and use these models
to predict the far-infrared emission. We find that our radiative
transfer model underestimates the observed FIR emission by a factor
two to three (see Fig.~{\ref{SED.pdf}}). Similar discrepancies have
been found for other edge-on spiral galaxies based on IRAS, ISO and
SCUBA data. Thanks to the good sampling of the SED at FIR wavelengths,
we can rule out an underestimation of the FIR emissivity as the cause
for this discrepancy. We argue that the most likely explanation for
this energy balance problem is that a sizable fraction of the
FIR/submm emission arises from additional dust that has a negligible
extinction on the bulk of the starlight, such as young stars deeply
embedded in dusty molecular clouds. The presence of compact dust
clumps can boost the FIR/submm emission of the dust while keeping the
extinction relatively unaltered (e.g. Silva et al. 1998; Bianchi et
al.\ 2000; Bianchi 2008).  An indication that embedded star forming
clouds might be the solution to the case of UGC\,4574 is that the
discrepancy between our radiative transfer model and the observed FIR
SED is stronger at shorter than at longer wavelengths. This implies
that warmer dust (such as in star forming regions) is necessary to
bring the model in balance with the data.

\end{document}